# Artificial Intelligence and work: a critical review of recent research from the social sciences

Jean-Philippe Deranty (Macquarie University) and Thomas Corbin (Macquarie University)

**Abstract**

This review seeks to present a comprehensive picture of recent discussions in the social sciences of the anticipated impact of AI on the world of work. Issues covered include: technological unemployment, algorithmic management, platform work and the politics of AI work. The review identifies the major disciplinary and methodological perspectives on AI's impact on work, and the obstacles they face in making predictions. Two parameters influencing the development and deployment of AI in the economy are highlighted: the capitalist imperative and nationalistic pressures.

**Introduction:**

This article reviews recent literature on the likely impacts of artificial intelligence (AI) in the world of work. It is one outcome of a grant-funded project whose aim is to map out the arguments for and against the idea that work is "central" for individuals and communities (see Deranty 2021, and the online repository: onwork.edu.au). Arguments for and against the importance of work have a long history (Applebaum 1992, Komlosy 2018), and they gathered renewed urgency with the rise of capitalism. In the last two hundred years, each generation has wondered about work and its importance, in constantly evolving economic, social, political and technological conditions. Today, debates on the centrality of work are shaped to a significant extent by the impact that artificial intelligence and machine learning are expected to have on economies, on social structures, and for working people.

This wide context explains why this review is not conducted from a specific disciplinary stance and why it covers broad array of issues, from the transformation of tasks and jobs to macro-economic consequences, all the way to political impacts. Research is being rapidly produced in all of these areas, and there is no clear way to view the many methods, assumptions, or findings alongside each other. In this paper we offer a critical review of the recent literature, bringing together the disparate scholarship on AI in the world of work and highlighting the problematic assumptions driving the leading interpretations and predictions regarding the future of work. Without a doubt, this lack of a specific disciplinary perspective and this broad scope have drawbacks. One can plausibly argue that only through particular social-scientific methods can specific features of an economic and social phenomenon be accurately described. And a broad scope brings with it risks of missing important references in each discipline. However, there might be benefits to taking an encompassing, non-specialised approach which outweigh these concerns. Such an approach might provide a more comprehensive snapshot of existing knowledge on the impact of AI in the world of work. Empirical research about platform work (Tubaro, Casilli et al. 2020, Casilli 2019, Tubaro and Casilli 2019) illustrates how difficult it is to keep in view all aspects of it at once. Even actors who are directly involved overlook aspects of the process as they consider it from their particular perspectives, based on their own interests and assumptions. A similar one-sidedness might affect specialist research. Studies of AI's impacts, its expected benefits and harms, are carried out by researchers in many disciplines (computer science, business, economics, management, organization studies, sociology, industrial relations, labour economics, history of economics and of technology, applied ethics, and more), using their disciplines' particular methodologies. Whilst specialised research gives access to particular aspects of the complex reality of AI, it is also important to have a view of the whole. The approach we take to the existing literature attempts to highlight the viewpoints from which assessments are made, the potential limitations built into these viewpoints. It suggests connections between aspects that tend to be looked at separately in specialist accounts. In order for such connections to become visible, it is necessary to be reflective of the context in which AI is deployed, as powerful background imperatives influence its development and deployment.

In section 1, developing these points about assumptions and scope in the study of AI, we define more precisely what we mean by "critical" in attempting a critical review. In section 2, we focus on issues of technological unemployment. Section 3 is about algorithmic management. Section 4 is dedicated to platform work. Section 5 considers the political dimensions of AI's impact on work.

## 1. Methodological considerations

### a. Definitions and scope

For the purposes of this study, we do not engage with the thorny issue of defining "intelligence" in Artificial Intelligence (see Wang 2019 for a thorough discussion). Our concern is simply with what AI-based work processes can achieve, what tasks AI can fulfill with, or instead of, human agents. More specifically, we are concerned with what AI can do in the framework of how economic activity is currently understood and organised. Work is a similarly difficult term to define (Budd 2011). For our purposes, however, work denotes the activities individuals engage in as part of the production of goods and services, for a profit if they are business owners, for a wage if they are employees, in the commercial, the public or the "third" sector. Such a vision of work is too restrictive, and we have ourselves criticized such a narrow take on work in other debates. Work covers activities of social reproduction, many of which are not counted in current economic classifications, and AI will be involved in these activities as well (Deranty 2021). However, to keep the study within reasonable limits, we use work here in the traditional, restricted sense of formal economic activity.

By AI's impact on work, therefore, we understand the computational methods that rely on the gathering and processing of data to replicate the activities human agents engage in as part of labour processes in the formal economy. These uses of AI include: coordinating machines and industrial processes (in manufacturing); managing all aspects of the workforce (HR, management, WHS); gathering, processing and evaluating information about business activities (accounting, forecasting, investing); predicting and evaluating outcomes for customers (expert advice in legal, medical and psychological matters); evaluating risks and benefits for customers and for internal purposes (insurance and finance); communicating with clients (customer service and counselling); anticipating, creating and managing customer needs and market demand (marketing and advertising); stocking and distributing material goods (logistics), including by transporting them (self-driving trucks and cars); supporting and potentially undertaking theoretical and applied research.

### b. A critical approach

The approach we are taking to the vast and diverse literature on AI's impact on work is "critical" in a sense captured by what might be called the Macbeth question: "Say from whence you owe this strange intelligence" (Act 1, scene 3). The core assumption this question captures is that scientific inquiries are inextricably tied to the historical context in which they occur. That context harbours specific tensions between human groups, conflicts between needs, views and interests, which lead to implicit and declared social-political conflicts. These tensions and conflicts mobilise all kinds of knowledges and even define particular epistemological standpoints. Whether they are aware of it or not, knowledge claims reflect social and political fault lines.

This means that AI did not emerge and is not being deployed in an economic or a political void. Digital innovation, like all modern technology, certainly has its own momentum, yet it is not sufficient to refer to advances in theoretical knowledge and technical improvements alone to explain the actual paths it travels and the forms it takes. The social context plays a direct part in the lines of technological development, both regarding the concrete features that artefacts and processes take, which objects and artefacts are produced and deployed in the first place, and the ways in which objects, processes and networks are put to use. This insight is well established of course. It has been validated and explored at length in the philosophy of technology (Feenberg 1991, Feenberg 2002, Feenberg 2012 for instance; Wajcman 2017 from a sociology perspective).

In studying the impact of AI on work then, it is crucial to keep in mind what kinds of imperatives and pressures it is AI innovation and deployment are under. AI processes are deployed in a context characterised by two fairly uncontroversial features: capitalistic imperatives and nationalistic pressures. These imperatives and pressures influence the paths taken by digital innovation and the forms in which it is deployed.

c. The capitalist imperative

A passage from Max Weber cited by Shoshana Zuboff in her introduction to *Surveillance Capitalism* (Zuboff 2019, 22) makes the point succinctly:

> "The fact that what is called the technological development of modern times has been so largely oriented economically to profit-making is one of the fundamental facts of the history of technology."

As Zuboff adds,

> "In a modern capitalist society, technology was, is, and always will be an expression of the economic objectives that direct it into action."

This applies in particular to AI and machine learning (Pasquale 2015 for an impressive account). Digital innovation has been driven to a significant extent by the attempts at winning the economic competition and increasing profit through the usual methods that are used in capitalistic economics (Brynjolfsson, E., & McAfee, A. 2014 for a candid version; Steinhoff 2021 for a critical, Marx-inspired take). Countless articles and books by business specialists describe the capitalistic advantage in firms investing in AI, with reference to positively connoted concepts like innovation, flexibility, adaptability and so on (for example Daugherty and Wilson 2018). Behind these terms are mundane economic mechanisms. AI has benefited from investment by companies that hope to find in it a new avenue for cutting costs, notably labour (Bessen et al. 2018) and transaction costs (Gurkaynak, 2019; Lobel 2018), increasing outputs through rationalization of the production process (Acemoglu and Restrepo 2019), raising productivity (Brynjolfsson, Rock, and Syverson 2019), managing the workforce in more efficient ways (Eubanks 2022), including through increased surveillance and control (Bales and Stone 2020), refining customer knowledge, and deliberately seeking to establish monopoly positions (Coveri, Cozza and Guarascio 2021; Rikap 2021). AI is viewed as a new way of reducing the labour share of income (Gries and Naudé 2018). As the core technology of platforms and the new business model they incarnate (Srnicek and Williams 2016), AI is seen by advocates of capitalism as ushering in a new, more agile and productive iteration of the system.

d. Nationalistic pressures

The second obvious feature of the current context determining the shape of AI is the nationalistic one. A major driver of AI innovation and development since the 1980s in the US has been the military (Berman 1992; Morgan et al., 2020 on US investment in AI for intelligence and surveilance systems). In more recent years, the battle for geopolitical hegemony has meant that other major powers, notably China, have also invested significantly in AI research (Barton, Woetzel et al. 2017, Allen 2019, Savage 2020). The geopolitical factor is directly tied to the economic one, AI development is stoked by an alliance of corporate and military interests: economic competition is one aspect of the battle for geopolitical hegemony, and military supremacy serves to ensure economic prosperity in competition with other nations (Hwang 2018 for the geopolitics of controlling semiconductors manufacturing and AI patents).

Underneath economic and military competition, an ideological battle is underfoot, one that pits the core values held by the different actors. To take one example, the Joint Artificial Intelligence Centre of the US Department of Defense presents its initiative as based on "solutions that are aligned with America's laws and values". There is a promise in the statement that AI innovations will deliver the kind of "good AI", or "AI for good", progressive-minded researchers and citizens are hoping for (Acemoglu 2021). But there is a more antagonistic aspect to such declarations, namely that the values embedded in the ethics of US-led AI will be American ethics, that is, a particularly American way of interpreting the core ethical norms that are to drive AI programs and algorithms. If we refer to the key norms of bioethics Floridi suggests we extend to AI ethics (Floridi and Cowls 2019), such as "autonomy", "justice", even "beneficence", these norms might mean different things in the Silicon Valley, in the Beijing technology district, and in European labs. The content of those values might well be irreconcilable. Beyond any cynical take on the power values might actually exert over the development of AI for world-powers and corporations fighting for hegemony, there might also be an ideological battle underway between, say, a liberal-capitalist, a social-democratic and a communist understanding of "good AI".

## 2. Technological Unemployment.

The first area of focus for studies of AI's impact on work is the threat of technological unemployment, an issue that has captivated imaginations for more than a decade. Debates about the impact of AI on employment revolve around its predicted quantitative impact, how much AI is likely to lead to machines replacing humans in work (2.2). This issue, however, depends on an understanding of the types of work activities AI is likely to perform, which in turn relies on assumptions about the skills involved in the tasks making up different types of jobs (2.1).

### 2.1 What types of jobs are affected?

AI is introduced in workplaces for increased efficiency in some technical aspect of the work process, or it is introduced explicitly with the aim of replacing human workers and thereby reduce labour costs. Whatever the reasons, a key condition of its success therefore is that it can replicate the outcomes achieved by human workers. Making this point does not commit one to "AI fallacy", the misleading belief that intelligent machines replace humans by replicating human skill use (Susskind and Susskind 2015). It might well be that machines achieve similar results as human agents via different mechanisms. It remains true however that successful outcomes in the labour process remain the baseline condition for the deployment of any workforce, human or digital. Predictions about technological unemployment rely on assumptions about the ways in which human workers achieve the outcomes expected in their jobs. The core concepts in discussions of technological unemployment therefore are those of skill, task, job, occupation and industry.

In much of the literature in the social sciences, notably in labour economics, management and the sociology of work, these terms are taken for granted. Discussions centre on the methods to accurately capture macro-economic trends and micro-economic issues, but not on the concepts themselves. This seems problematic though. To understand the impact of AI on work, one should not overlook the empirical complexity and conceptual slipperiness of a concept like "skill". Attewell (1990) and Spenner (1990) show well the range of possible meanings, and how the intuitive appeal of a notion like "unskilled" work in fact is anything but evident, how evaluation of "skillfulness" can shift depending on the perspective taken. Industrial sociologists demonstrate through grounded case studies that intuitive assumptions about "routine work" can be deceptive (Pfeiffer 2016). Researchers in education question the conceptual validity of the concept (Clarke and Winch 2006), particularly after it expanded with the shift from industrial to post-fordist frameworks where a whole array of communicative, social and emotional abilities, as well as personal attributes such as work commitment, were added to traditional craft knowledge (Payne 2000). Concrete issues arising from overlooking the complexity of skill will be discussed in the final part of this section.

It is well-established that previous waves of automation propelled by advances in information and communication technology were biased in favor of workers with higher skills, replacing lower-skill workers and assisting workers with pre-existing complex skill sets (Acemoglu and Autor 2011, Göranzon and Josefson 2012, Buera, Kaboski et al. 2015, Mellacher and Scheuer 2020). Classical work in labour economics on the differentiated impact of innovation (Tinbergen 1974) has led to sophisticated econometric models to formalize "skill-biased technological change", which provide economic descriptions of the premium that technological innovation gives workers with higher skill, in terms of wage increase and the very availability of jobs. The question is whether the established tenets of labour economics remain true for AI.

AI works by processing large amounts of data to identify patterns (Van Rijmenam 2019). It does this particularly well when there are set parameters to the data and set aims for the patterns. Because of this, AI technology is particularly efficient in task-oriented, routine environments where large amounts of data can be analysed to identify patterns, make decisions based on those pattens, and produce solutions or efficiency dividends (Neufeind, O'Reilly et al. 2018, De Vries, Gentile et al. 2020). For these reasons, there seems to be a case for interpreting the effect of AI along the same lines as the previous wave of automation through computerisation. That is to say, AI will replace large numbers of jobs and routine work which is often manually conducted and which requires low expertise. There are several factors, however, that complicate this picture significantly.

First, a detailed look at what professionals actually challenges the assumption that work that appears more complex or requiring higher skills is necessarily equivalent with non routine work. A significant portion of professional work in fact involves routine activities (Ford 2015, 2021; Susskin and Susskin 2015; 2020) such that, even if some "higher" cognitive components are involved (memorisation, or complex judgement, or evaluation), "higher skill" jobs are themselves open to automation by AI.

Secondly, the most striking aspect of AI-based automation is the capacity of machines to operate autonomously, to "learn" rather than function solely on preset patterns. As a result, labour economists, notably Acemoglu, Autor and Restrepo, highlight AI's relationship to "high skill automation" (Acemoglu and Restrepo 2018), which compounds the exposure to automation of "higher skill" jobs. The Susskinds' related prediction of "the end of professions" is corroborated by a number of reports (for instance Manyika, Chui, and Miremadi 2017).

Thirdly, many entry level and manual jobs are in fact not routine and therefore not easily codifiable (Goos, Manning et al. 2014, Autor, Dorn et al. 2015, Barbieri, Mussida et al. 2020). Some basic -skill jobs are thus uniquely immune. This rests upon the famous Moravec Paradox which "refers to the striking fact that high-level reasoning requires very little computation, while low-level sensorimotor skills require enormous computational resources." (Van de Gevel and Noussair 2013). Some skills which come naturally to human beings require massive amounts of computational power to replicate and consequently, "it will be hardest for new technology to replace the tasks and jobs that workers in the lower-skill level occupations perform, such as security staff, cleaners, gardeners, receptionists, chefs, and the like" (Gries and Naudé 2018).

Finally, human skills can be complemented rather than copied by automated processes, with machines taking charge of the routine aspects of the job, for instance in legal work (Brooks, Gherhes et al. 2020; Alarie et al. 2019). And human work can complement machine work, through "humanly extended automation" (Delfanti and Frey 2021). This is the case, for instance (Ebben 2020), when low-skill tasks continue to be fulfilled by humans, in the service of automated processes, for instance in warehouse work. In this case, human work continues to be performed by humans, not because it is difficult to automate, but because humans are better at it, or cheaper to employ, than machines.

## 2.2 Substituting, complementing, or creating human work

### 2.2.1 Pessimistic scenarios

Many economists and technology experts contend that AI will substitute for human work at such a scale that social-economic organisations will be shaken to the ground as a result. This is a major aspect of debates on the centrality of work today, often the initial argument for "post-work" models of social organisation (typically Danaher 2019).

Universally cited references are, first of all, Brynjolfsson & McAfee's publications, notably *Race against the Machine* (2011), and *The Second Machine Age* (2014). The two business and technology experts extol the capacity of intelligent machines to lift productivity, massively increase outputs and spur wealth creation, driving prices to zero for some commodities. Their celebrations of the digital revolution, however, come with warnings about the severe impact of AI-driven automation on labour markets, as technological advances create more losers than winners because of skill- and capital bias. The policy solutions they call for are premised on the dangers of automation, and of AI in particular. Another ubiquitous study is Frey and Osborne's 2017 *The Future of Employment: How susceptible are Jobs to Computerisation?* (2017) The study was cited over 10,000 times at the time of writing (Google Scholar, January 2022). The two business scholars famously predict that 47% of all jobs within the U.S. are at risk of technological replacement within two decades. Using a similar approach, Bowles arrived at an even higher figure, claiming that 54% of jobs in the EU and USA were under threat in the same time span (Bowles 2017). Since then, many studies, using a variety of methods have added to these anticipations (Halal et al. 2017; Schwab 2017; Chessell 2018; Gruetzemacher, Paradice et al. 2020; Gruetzemacher, Dorner et al. 2021). In a multi-country approach covering 32 nations, Nedelkoska and Quintini (2018) estimate that 14% of jobs are highly automatable (probability of automation over 70%), 32% have a risk of between 50 and 70%. The figures are confirmed by Pouliakas (2018) via a method that uses disaggregated job descriptions in a key survey conducted by the European Union (the European Skills and Jobs Survey, covering 49,000 EU adult workers), a method that allows him to factor in information on skill requirements (for China, Zhou et al. 2019).

Beside displacement, another trend widely anticipated is the polarisation of labour markets, similar to what occurred in the previous wave of automation (Autor, Katz, and Kearney 2008; Goos and Manning 2007; Autor and Dorn 2013; Michaels, Natraj, and Van Reenen 2014; Goos, Manning, and Salomons 2014; Graetz and Michaels 2015; Frey 2019; Autor, Dorn, and Hanson 2015; Scarpetta 2018; Bordot and Lorentz 2021).The reinstatement effect (see next section) might favour only workers with specialised skills (Holm and Lorenz 2021), whilst new jobs might be created in occupations that are "technologically lagging", where automation cannot enter for price reasons, but lower skill attract lower wages and precarious conditions (Petropoulos 2018). Consequently, AI might lead to a hollowing out of white-collar jobs in business, administration and knowledge-industries. These concerns are confirmed by a European Parliamentary study (Deshpande et al. 2021).

2.2.1. Optimistic scenarios

Many economists, historians, business scholars and executives reject such pessimistic visions of massive technological unemployment.

Large surveys of employers present contrasting evidence. A key Manpower survey in 2017 provided interesting figures, with managers in some countries expecting to substitute workers for machines, whilst others expected AI to increase hiring. Overall, the survey demonstrates optimism (also ServiceNow 2017). Via a questionnaire targeting 3000 companies Bughin (2020) concludes that labor redistribution will occur.

Global accounting firms and business consultancy groups are often adamant regarding the potential for AI to increase productivity. Purdy and Daugherty in a 2016 Accenture report estimate that AI has the potential to increase labor productivity by up to 40% in 2035. Gillham et al. in a PWC report estimate that global GDP would increase by 14% by 2030, an equivalent of up to $15.7 trillion. All geographical regions of the global economy are said to benefit.

Some current empirical work supports this, with studies published by international institutions reporting the absence of any current impact on job markets (Georgieff and Milanez 2021 for the OECD). The ILO has published many reports on the future of work, which tend to be cautiously optimistic (for instance 2018). A 2018 study by the World Economic Forum (2018) predicts that automation will result in a net increase of 58 million jobs, with a total 133 million new roles created and 75 million current workers displaced. In a recent OECD paper, Squicciarini and Staccioli (2022) study the impact of natural language processing techniques on specific occupations and find no significant effect on employment. A study by Acemoglu, Autor, Hazell, and Restrepo (2020) concludes no significant effect yet in the last decade at the occupation or industry level in the US. This confirms Autor's work of the early 2010s, in which he cautioned against overly pessimistic conclusions about technological unemployment (2015; Autor and Handel, 2013).

The standard approach to discuss the likelihood of technological unemployment is by labour economists who devise methods to extrapolate the impact of technology on particular tasks and deduce from it the impact on occupations and industries more generally. Using this approach, Frey and Osborne's earth-shaking predictions on the impact of AI was refuted by Arntz, Gregory and Zierahn (2016). In their report for the OECD, the German labour economists modified Frey and Osborne's approach by employing an alternative method for linking tasks to occupations (see also their discussion paper 2019), With this new method, they found evidence for much lower outcomes, around 9% across the OECD (a high of 12% in Austria) and as low as 6% in South Korea. In another noteworthy approach, Princeton computer expert Edward Felten and his colleagues developed a model for how computerisation and specific AI functions can affect the activities making up particular occupations. They also land on more measured conclusions (Felten, Raj, Seamans 2018. Felten et al. 2019, Felten, Raj et al. 2021).

One approach that can be used on its own but is also often combined with labour economics modelling, draws on evidence from economic history. The history of automation to date demonstrates that new technologies so far have consistently created new jobs, both directly and indirectly (David 1990). In an important review of the literature, which combines a historical approach with a task-focused one, Ekkehardt et al. (2019) conclude that AI in fact is likely to play out in ways different from previous waves, by increasing productivity and potentially creating more inclusive growth, provided correct educational measures are taken in countries that are part of globalised labour chains.

At the heart of the debate are arguments from mainstream economic theory. AI is viewed by many business experts as a technology that can spur innovation and provide competitive advantage (Davenport 2018). Innovation brings with it a "productivity effect": increased productivity in one sector raises labour demand in other sectors. As Smith (2020) explains for instance, "the automation of one industry means higher demand for labor in other industries like the production of machines, the cultivation, extraction, or processing of raw materials, and the building of infrastructure like ports and highways."

The productivity effect plays out in complex ways. Amongst the many economists to have studied it, the most influential ones in current debates on AI-driven automation are Acemoglu, Autor and Restrepo, and the authors they have collaborated with. In their contribution to *The Economics of Artificial Intelligence* (2017) Acemoglu and Restrepo summarise in plain terms the complex logics associated with the productivity effect. First, as noted, there is a rise in the demand for labor that follows automatically from economic growth triggered by innovation. In theory the demand for labour can be witnessed even in sectors where automation occurred. Second, demand for labour can increase because automation triggers increased demand for capital. Third, automation can deepen existing automation, which increases productivity without substituting labour (since it is machines that are improved). And fourthly, and most importantly, AI-automation creates new tasks. The creation of new tasks is a key component of the labour economists' arguments against universal technological unemployment.

Another macro-economic argument highlighted by Autor (2015) is captured in the image of the O-ring. It is worth citing a passage at length, as it illustrates many of the points raised by economists:

> "tasks that cannot be substituted by automation are generally complemented by it. Most work processes draw upon a multifaceted set of inputs: labor and capital; brains and brawn; creativity and rote repetition; technical mastery and intuitive judgment; perspiration and inspiration; adherence to rules and judicious application of discretion. Typically, these inputs each play essential roles; that is, improvements in

> one do not obviate the need for the other. If so, productivity improvements in one set of tasks almost necessarily increase the economic value of the remaining tasks. An iconic representation of this idea is found in the O-ring production function studied by Kremer (1993). In the O-ring model, failure of any one step in the chain of production leads the entire production process to fail. Conversely, improvements in the reliability of any given link increase the value of improvements in all of the others. […] Analogously, when automation or computerization makes some steps in a work process more reliable, cheaper, or faster, this increases the value of the remaining human links in the production chain." (Autor, 2015)

One specific dimension of the O-ring mechanism is that, by allowing for increased automation in the industrial and manufacturing sector, AI might have a "multiplier effect" in service occupations connected to them. On some accounts, this effect might even be felt in manufacturing industries servicing automated factories (Goos, M. et al. 2015).

One other argument combining economic and historical dimensions relates to the specificity of productivity increase resulting from AI. It is a well-known fact that productivity has been stagnant in developed economies over the past seventy years, with growth per decade decreasing from 2.3% in the 1950s, to 1.8% in the 2010s (Gries and Naudé 2018). Lewis (2018) has shown that labour productivity growth in the UK since 2007 was the lowest decade on average since the 18$^{th}$ century. Even in the last decade, productivity growth slowed significantly (Brynjolfsson, Rock et al. 2019). As early as 1987, economist Robert Solow famously quipped: "you can see the computer age everywhere except for in the productivity statistics". Some researchers claim that AI may be the key to reversing this trend. As Munoz and Naqvi (2018) write, "the world is seeing a solution to its productivity woes and the answer lies in the rise of Artificial Intelligence." Leading economic historian Joel Mokyr agrees (2018) and is resolutely optimistic about AI's potential to restart growth. In response to the objection that increase in the productivity cannot be viewed in statistics today, Brynjolfsson et al. (2017) argue that technological innovation suffers from an "implementation lag" and so accurate measurements of the impact of AI on productivity are yet to be known.

Another dimension of accounts defying pessimistic and dystopian scenarios relates to the kind of jobs that AI might make possible, with the claim that it might make many jobs, either existing ones or new ones, more satisfying, as they might involve higher skills or more creativity from the workers (Makridakis 2017). AI processes might also improve working conditions. For many, AI "can reduce the risk of dangerous or unhealthy working conditions, encourage the development of specialist or soft skills, and improve accessibility to certain jobs" (Depschande et al. 2021).

2.2.3 Critical assessments

a.  Methodological doubts

Predictive exercises must meet formidable methodological challenges. The example of labour economics is informative. In order to assess the trajectory of labour market, labour economists and computer experts first make lists of skills and abilities associated to particular occupations, which they gather from the datasets of national labour offices and other organisations (typically the Burning Glass Labor Insight or the O*Net in the US), or from online job vacancy listings (Acemoglu et al., 2020). They then use statistical tools to connect these lists with what AI processes are assumed to be able to perform. The competitive tipping point at which AI becomes financially attractive and thus substitutes for labour is calculated via established models of neo-classical economics. The new models these exercises produce are standardly tested against established employment trends that responded to previous technological innovation. For the external observer, these methods seem to deliver substantial lessons for understanding past trends, but are far less convincing when it comes to future ones. The frequency with which the mathematical models are revised, with new parameters and axiomatic hypotheses being introduced in each new paper, gives the non-specialist the sense that there is a gap between the assurance with which conclusions are stated in plain English and the ability of the models to capture reality, notably given the number of idealizing assumptions underpinning the analyses. The models appear to describe mathematically consistent worlds, but it is less clear that they describe our messy one, let alone what it might be in the future. At the macro-level, the factors involved in economic reality across different context are so numerous and variable, one would assume significant unpredictability in how AI might complement rather than replace human work. As Acemoglu, Autor and Restrepo themselves have shown, the complementary effect leads to the creation of new tasks, notably as worker's time is liberated from routine, existing jobs take on new content, and new needs for new tasks arise. New technology creates entirely new jobs (Wilson, Daugherty et al. 2017). By definition, the lay reader is tempted to say, if needs, tasks and jobs will be new, it seems difficult to guess what they might be, let alone capture them in models premised on existing job profiles, and on the descriptions of the tasks entailed in currently existing occupations (see the 2018 ILO literature review for precisely this point). The predictive challenge is compounded by the slipperiness of skills and tasks noted above, at ground level so to speak. Methods for making predictions on the impact of AI on tasks by using large datasets seem ill-suited for capturing the complexity at the micro-level of what particular jobs actually involve in their specific context (Ebben 2020). This is true for technical reasons, because many jobs are actually more difficult to perform than is often assumed, but also for economic ones: as De Stefano makes the case (2020), assuming that automation necessarily increases productivity might in some cases be based on an overly narrow view of the tasks automated and on erroneous measures of efficiency.

Similarly, methods that rely on case studies only inform on a particular job or occupation at a particular time and place, and there are methodological risks in generalising from particular cases. Many researchers highlight the difficulty of generalising from one national context to another, as different structures of the economy, the education system, and so on, mean that AI impacts work differently in different economies (Spencer and Slater 2020). In a journal that has published key studies in this area, an important review by Clifton, Glasmeier and Gray (2020) highlights that, "the impact of technology on employment is not deterministic—the deployment of these new technologies is contingent upon a multitude of factors, including public policy, firm strategy and geography, among others."

Finally, external conditions directly impact on the deployment of AI. As Hwang shows in his landmark study (2018), even though they are taken for granted in most reports, computational power supported by adequate hardware (notably quality microconductors), and energy availability, are basic material conditions of AI systems. This makes them non-trivial conditions to be taken into account when calculating the likelihood of human tasks and jobs being replicated in the real world. Geopolitical, economic, resource limitations might well slow down or hamper AI deployment simply because material support is lacking. Similarly, in a world prone to climate crisis, the environmental impact of AI (Strubhell et al.2019; van Wynsberghe 2021) might well constrain its deployment, at least if sustainability becomes a serious parameter in economic analysis and economic activity.

b. Marxist critiques

A number of authors consider AI specifically as a form of capitalistic innovation, and show that the logic of capitalism dictates that work is a long way from becoming obsolete. The absence of technological unemployment is in itself nothing to celebrate though, as it coincides with higher levels of precariousness, underemployment and exploitation. Drawing on Cohen's classical reconstruction of historical materialism (2000), Barbara Nieswandt (2021) shows that private property in the means of production and the profit motive, two structural features of capitalist economies, make it unlikely that AI-based automation will lead to massive job losses. Strict property rules mean that technological innovation is exclusively owned, put to use, and the outcomes of its productivity-raising potentials captured, by private owners. The capitalist imperative means that the deployment of technologies in a capitalist context is guided exclusively by the search for profit. The combination of these two factors means that technological innovation in a capitalist economy serves to increase output as a way to increase profit. Other alternatives cannot be countenanced given the rules of the game. This is true not just of the altruistic alternative that would mobilise technology to reduce working hours, but also of the possibility of using productivity to reduce the wage bill whilst keeping production constant. Given the other tools capitalists possess to ensure the exploitation of workers, the search for profit is better served by increasing output than job churn, which means that there is no incentive to shed jobs. As Dinerstein et al. concur (2021), the decisive factor to consider is "not technological opportunity" but the "profitability criterion".

Another argument draws on readings of Marx that emphasise class antagonism as a key explanatory factor in the organization of capitalist production, including in its adoption of technological innovation. A major use of technology is to bypass labour forces when they are well organized and push back effectively against capital imperatives (Mueller 2021). When the capacity of labour to organize is weakened, the need to shift work to capital to bypass labour is less pressing. This argument is confirmed by Fleming through his focus on power in organisations (Fleming 2018). Labour's current deficiency in power compared to capital's means that automation is "bounded". This is compounded by the cheap cost of labour, itself an effect of weakened labour protections. When human labour is relatively cheap, there is no incentive for capital to invest and deploy expensive technologies. Surplus-value can be extracted just as well from human workers (Dinerstein 2021).

Thirdly, in the current phase of capitalism, with its distinctive property structures and ideological underpining, it is shareholder value, not productivity that matters. If shareholder value can be ensured through other means than investment in technology, which comes with significant sunk costs, then that path will be chosen. With the sophistication of financial tools and the protection of promiscuous taxation schemes, there is no pressing incentive in many industries to invest in technology designed to replace human labour (Smith 2020). In other words, even if the previous arguments did not obtain, it would still be the case that the current economic context does not incentivize massive deployment of job-replacing technologies.

A fourth argument (Benanav 2020, Smith 2020) concentrates on the differential impact of productivity across sectors of the economy. Overcapacity in manufacturing and agriculture leads to an exodus of workers towards service and care sectors, where productivity is achieved through wage suppression and depreciation of working conditions. Already in the early 1990s Gorz predicted that automation would result in the rise of new occupations centering on personal services, to tend to the needs of an elite of technical and knowledge workers (Gorz 2011). This triggers employment in new service sectors, even a return to older forms of dependent labour. These new occupations come with low wages, precarious working conditions and tenure and uncertain hours. But there is no massive job churn as a result of massive deployment of automated work processes.

These Marx-inspired analyses of technological unemployment are supported by other analyses of historians who directly contradict optimistic readings like Mokyr's, and emphasise the slowing down of innovation under financial capitalism, where profit-maximisation occurs through speculation rather than changes in industrial paradigms (Gordon 2014, 2015).

3. **Algorithmic Management**

In this section, we shift from macro- to micro-issues of work, where AI is already having an impact.

Algorithmic management covers the tasks traditionally performed by human managers: the hiring of employees (from CV selection to automation of the hiring process), optimisation of the labour process (through the tracking of worker movements, for instance GPS tracking or route-maximisation in transport and logistics), evaluation of workers (through rating systems), automated scheduling of shifts, coordinating customer demand with service providers, monitoring of workers behaviour, algorithmic incentivisation (through algorithm-based "nudges" and penalties) (Duggan et al. 2020 for a thorough review). Business scholars highlight the technology's ability to improve work flows, for instance for optimal job allocation (Jarrahi et al. 2021), to cut costs, say in hiring, and to improve predictive power in all dimension of the business activity. From this point of view, AI based algorithmic management offers organisations the chance to delegate decision making power to more efficient and effective managers (Von Krogh 2018).

AI needs data regarding workers' skill, time use, and behavior (Nguyen, Aiha 2021), which in turn makes worker monitoring a necessity. The more data is fed into AI processes, the more effective its use (Gal, Jensen, & Stein, 2020; Ebert, Wildhaber, & Adams-Prassl 2021). Some aspects of worker monitoring seem benign and might even be benevolent, as when it is used to increase digital security, prevent fraud, or to monitor and improve worker health and safety (De Stefano 2019). However, many critical management and organizational theorists, labour lawyers and sociologists of work, as well as scholars studying human-machine interactions, highlight concerns with the spread of algorithmic management. AI further increases the power-imbalance between managers and employees (Jarrahi, Newlands et al. 2021)., notably as a result of the asymmetry of information.In a lucid report from the Data and Society Research Institute, Mateescu and Nguyen (2019), usefully summarise these concerns around four main points.

Surveillance and control: algorithmic management raises obvious issues of privacy (Bave, Teo, Ladal 2020; Ebert, Wildhaber, Adams-Prassl 2021), not just at the workplace, but also at home, notably following the pandemic-induced shift to home-based working (Collins 2020). Privacy infringements can occur at all stages of the data cycle: at the time of collection, in the analysis of the data, in the use of the data, and when data ought to be erased. Breaches of privacy touch on a fundamental human right, but they also represent a strong leverage tool for managers, with which they can exert control and undermine autonomy (Shapiro 2018). Surveillance can also lead to increased pressure on workers to perform, taking away moments of respite, as is well documented in warehouse (Hanley and Hubbard 2020) and platform work (Newlands 2021). This can have severe and long-term impact on well-being. Algorithmic control of the work process takes away the dimensions of personal intervention, choice and even of creativity (Huang 2021).

A key study of algorithmic management, emphasising the contestation between management and employee around the new tools of control and coercion offered by AI, is Kellogg (2020), which uses labour-process theory as its framework (Gandini 2019 for gig work ). This framework is particularly apt for studying the concrete ways in which the capitalist imperative translates into the attempt by management to control the workforce at the point of production. Kellogg is particularly useful for its survey of literature making the case for algorithmic management as a new tool for increased efficiency in the running of organizations, through better decision-making, better coordination and better organizational learning.

Algorithmic management might perpetuate societal biases and reproduce discriminatory practices at work, whether the discrimination is built into the algorithms, or management's use of the algorithm, or as a result of customer's rating of workers (Noble, 2018; Benjamin, 2019; Kellogg, Valentine et al. 2020; Akter, McCarthy et al. 2021)). Increasing amounts of empirical evidence confirm this risk (Obermeyer, Powers et al. 2019; Datta et al., 2015; Lambrecht and Tucker 2019). AI also gathers and processes data in ways that are often hidden from workers, leading to decisions made in ways which humans cannot, or at least cannot quickly or efficiently, process. Indeed, if AI processes could be understood and tracked transparently, this would undermine the point of having them in the first place (Buchanan and Badham 2020). Algorithm-based decisions emerge as from a "black box" (Pasquale 2015) or "magic box" (Thomas, Nafus et al. 2018). Attempting to understand and track AI processes would undermine the point of having them in the first place (Buchanan and Badham 2020). Algorithmic management therefore makes a unique demand for the "blind trust" of workers (Leicht-Deobald, Busch et al. 2019). Workers are managed and directed through reasons they are not provided with, preventing collaborative or even consultative leadership styles, and instead substituting a directive or even coercive one in its place (Dunphy and Stace 1993).

Finally, the literature raises accountability concerns. Management by AI is by definition about removing the human element from the decision-making process. A number of Human Relations Management experts view this with skepticism (Duggan et al., 2020). The machinic objectivity of AI-based management processes has the effect of creating a screen between worker and organisation, and between worker and management. The company appears to be absolved of its responsibilities, removing existing avenues through which workers understand and demand these responsibilities are met. Workers are put at the mercy of processes they have little control or recourse over (Loi, Ferrario et al. 2020, Veen et al., 2020, Purcell and Brook 2020, Joyce and Stuart 2021).

4. **Platform Work**

The power of AI for gathering and processing vast amounts of data can be harnessed in traditional work settings where employees are hired under work arrangements and labour contracts predating AI automation. One of the interesting lessons of Srnicek' 2017 *Platform Capitalism* is to draw attention to industrial platforms operating in factory settings, pointing to an aspect of AI far less visible than its use in gig work. However, the computational power of AI allows it to function not just as a new industrial tool within preexisting work processes. It also becomes the centrepiece of a new business model that radically alters modes of working, as well as the conditions of employment and the interactions of workers with management and customers.

Since they first emerged in 2008-2009, AI-backed platforms have attracted a vast amount of scholarly attention. Analysis of their structure and functioning has been performed mostly by communication theorists and media specialists, and the analysis of the new modes of work and employment they generate by sociologists, management experts and organizational theorists.

One way to conceptualise platform work is by focusing on the relationship between worker, employer and customer. This was the approach taken by Duggan and colleagues in a 2017 review of literature on "gig work" published in a journal of Human Resources Management (also Schmidt 2017). This focus leads to a useful distinction between capital platform work, crowdwork and app-work. The first corresponds to what is known as the "sharing economy", where individuals use the platform as a digital commodity market to sell goods or assets (like usage of their accommodation). Workers here operate like small entrepreneurs. In crowdwork the platform is like a digital labour market, where jobs are tendered to potential workers and the work is be performed via the platform (as in Google's Mechanical Turk). There are different types of crowd-work, whether large tasks are divided into small tasks performed by different individuals, or similar work is performed simultaneously by several individuals. In crowdwork, the relationship of worker to employer is minimal. App-work is work provided in the physical world (food delivery, ride-hailing services) where the platforms connects worker and customer. This is the type of platform work that has attracted the most attention, notably because management issues are prominent.

In order to find some bearings in the large literature dedicated to platform work, one of the most useful resources to consult is the review by leading sociologists of work Steven Vallas and Juliet Schor (2020). Their sociological lens invites them to construct a taxonomy based not on employment relations but on the groups performing different types of work (notably along a scale of complexity of skills). This taxonomy offers another entry point to study conditions of platform work (income levels, geographical location, terms and conditions) and the issues each group of worker encounters specifically. The sociologists identify five groups of platform workers: architects and designers of platforms; cloud-based consultants; gig workers; micro-task crowdworkers; and content producers who perform "aspirational labour".

The study proposes a map of the rich terrain of recent studies of platform work. Four thematic families are identified and critically discussed.

The first is the utopian view of platforms as an instrument enabling individuals to share goods and services outside the corporate form, where work evades the traditional management interaction, a form of production and exchange which some authors think will boost economic activity and create a new form of capitalism by reducing transaction costs, and as the peer-to-peer interaction fosters trust. A key reference here is Sundararajan 2016. Vallas and Schor reference a large number of studies on working conditions for app-workers (see the more recent Moore and Woodcock 2021), and conclude that power rather than horizontal democracy has so far been the experience of workers.

The second perspective is the polar opposite, raised by researchers who see in platforms a new form of Weberian "iron cage". Rahman' (2021) focus on "invisible cages", for instance, explores the experiences of freelance platform workers contending with opaque algorithms evaluating their performance and dictating their future success. Here, the emphasis is on the different forms of control platforms enjoy in comparison with traditional workplaces, notably through surveillance, monitoring, but also the gamification of work, symbolic rewards and inducements (see Perrig in Moore and Woodstock 2021). As Vallas and Schor (2020, 278) write, "Max Weber's fears regarding bureaucratic subordination (the iron cage, however translated) pale in comparison with the prodigious powers over human labor that digital technologies are thought to enjoy." However, this view of platforms underestimates the capacity of platform workers to evade control, resist and organise.

The third image of platforms sees in them an economic model that only accelerates a process of precarisation that was already under way under previous regimes. Vallas' own work, some of which in collaboration with Arne Kalleberg, another leading sociologist of work, is one major reference here (2017, 2018). Vallas and Schor reject an assumed view of platform workers that is overly homogeneous. Many platform workers in fact use the work to complement income, precarisation through platform is far from a universal trend.

The fourth family of studies focuses on platform technology's ability to be put to very different uses depending on the institutional context. Platforms can create new forms of control, or instead help to regulate the work and protect the workers. Against, this, the two sociologists point out that there are fixed attributes to platform work, that resist institutional shaping, indeed that the most powerful platforms are the ones that shape their institutional environment.

Vallas and Schor themselves present an alternative image, one where power is centralised for the key technical and economic functions, but control is distributed and largely relaxed. Platforms, they argue, provide a new mode of governance and model of economic activity. Most importantly, "platforms greatly relax personnel selection criteria and affords workers considerable autonomy over when and how often to work (283)". One feature of this relaxation of management control, is the heterogeneity of the workforce this produces, which acts against orgnanisation. Whilst AI can be used in traditional settings for increased surveillance and control, on platforms, the two sociologists argue, surveillance cannot be as strict. The heterogeneity of the workforce exists also in the locations of the workers, who are scattered around regions, and even around the world. This creates isolation and again prevents organisation for collecive action.

One aspect of crowdwork that is worth highlighting is micro-tasking crowdwork that goes to the heart of AI itself. As sociologists of technological innovation, a lot of human labour is needed to fill the cracks of AI (Gray M, Suri S, Ali S, et al. 2016; Tubaro P and Casilli A 2019; Tubaro P, Casilli A and Coville M 2020). This human labour is often obscured as "corporate communication highlights the role of technology, not human contribution, especially in the AI industry" (Tubaro 2021; along similar lines, Newlands 2021). Gray and Suri (2019) term this human work invisible to the outside but necessary for "automated" processes to function, "ghost work".

A key lesson from the sociology of work perspective is that it is impossible to make overly general statements about platform work. Different types of workers operate under different conditions and have vastly different experiences. This is true for instance of job quality. Micro-tasking crowdwork can be hugely injurious on emotional and mental health (Hermosillo et al., 2021), for instance for workers screening violent content. A lot of crowdwork and gigwork is highly flexible, has non regular hours, which can have a negative impact on work-life balance and affect physical and mental health and personal relations as a result (Muntaner 2018, Allan et al., 2021). The feature Vallas and Schor highlight most is the devolution of control in many platform environments, allowing workers to enjoy more autonomy in the completion of their tasks and the relationship with clients. Service work for an Uber driver for instance is far less scripted than service work in more traditional settings. Flexible hours might also be viewed as increased autonomy. But what can in some conditions amount to an increase in autonomy can also mean a lack of training, notably around OHS issues. Workers are left to their own devices in many gig environments. By the same token, however, AI can also complement work in some professions, notably by taking away the routine aspects of the job. Workers then focus on the more creative, or rewarding parts of the work.

## 5. Conclusion: the Politics of AI work

Amongst the many potential societal impacts of AI, this review focuses on those that affect the world of work. If, as a result of AI deployment in work, technological unemployment does occur at a significant scale, or wealth polarisation, or crowdwork and app-work become widespread employment models (so far they are not), bringing further precariousness in employment conditions, then the impact on current social organisations will be significant. This is because current social systems continue to be organised on the basis of full-time employment in jobs attracting good wages as the condition for the complete enjoyment of social and economic rights, including full health and pension coverage. One of the consequences of the deployment of AI could be further stress being put on already ailing systems of social protection (Konkolevsky 2017).

In response to the possibility of social crisis resulting from the new industrial revolution, many progressive thinkers advocate turning this threat into an opportunity for a radical transformation of our thinking on social and economic life, through some version of a basic income (Susskind 2021), leaving behind the modern work ethic (Weeks 2011), indeed harnessing full automation for a leap into "luxury communism" (Bastani 2020). Outside this blue-sky literature, we find a spectrum of normative answers to AI challenges. At one end are philosophical studies of the "ethics" of AI, which tend to overlook macro-economic and social contexts and focus instead on the ethical norms embedded in particular AI processes and machines (Floridi 2014 as an informative exception). Other social-scientific approaches, tend to combine descriptive and normative foci. Much of the normative discussion in these literatures is in the form of: "governments ought to do x, y, z", or "such and such regulatory principles ought to be enshrined in AI development" (Arogyaswamy, 2020). Typical in this respect is the last part of Susskind's *A World Without Work*. To counteract the deleterious effects of technological unemployment, Susskind canvasses as the solution a "big state". This is a state that taxes elite workers who manage to remain relevant in a depleted labour market, privileged individuals inheriting wealth, big business and what Marx called "constant capital" (machines); a state that introduces a conditional basic income; and shapes the ways in which individuals fill their leisure time to find meaning and purpose. As a blueprint, Susskind's proposal is detailed and seems consistent, but it is utterly unrealistic in current ideological and geopolitical conditions, particularly in the few countries spearheading AI innovation. Ideal-scenario policy recommendations might have their usefulness, but the immense gap between the propositions and economic and political reality points to the need for another kind of approach, one that focuses precisely on what makes these developments of AI doubtful, at least in the short term. Recent research in political science and labour law focuses on these aspects. What is at stake is the clash between the two imperatives noted at the outset and the plausibility of a "good AI", or "AI for good". Regarding the capitalist imperative, a number of arm wrestles are underway between a few immensely powerful corporations and the actors representing workers: activists, academics, trade unions, cities and regions affected by AI (typically by platforms such as Airbnb or Uber), branches of national governments, multinational organisations like the EU, international bodies, like the ILO. One major battle concerns the legal status of gig-workers, whether platforms can divest themselves of all responsibilities regarding the people they employ. For the US, the research of labour law expert Veena Dubal is particularly significant (for instance Dubal 2021). Another battle is about the right to privacy of workers (De Stefano 2020 for Europe). Another battle concerns the development of regulatory frameworks to give an ethical frame to AI development. AI companies, notably the largest platforms, are actively counteracting these efforts, through direct and indirect political interventions, intervening in formal legislative processes through lobbying, challenging legal decisions, mobilising their customer base (Schor and Vallas 2019; Thelen, 2018; Collier et al. 2019), or even directly flexing their digital muscle against cities, regions and even nation states (such as Australia in 2021, Quinn 2021).

Regarding the nationalist imperative, one might be sceptical of the willingness of states to enforce principles of "good AI" when they are engaged in high stakes contests over geopolitical hegemony. For instance, one aspect of the struggle concerns the control of the supply chains and production factors behind the manufacturing of semiconductors (Hwan 2018). Given the rhetoric used by the main actors, it is difficult to see ethical considerations having much sway for some of the possible extensions of AI.

What these extraneous, economic and political, parameters indicate is that AI does not by itself determine 'good' nor 'bad' outcomes for the world of work. Rather, what matters is the kind of world in which AI will be developed and deployed.